%% file: templateArxiv.tex
\documentclass{article}

\usepackage{PRIMEarxiv}

\usepackage[utf8]{inputenc} % allow utf-8 input
\usepackage[T1]{fontenc}    % use 8-bit T1 fonts
\usepackage{hyperref}       % hyperlinks
\usepackage{url}            % simple URL typesetting
\usepackage{booktabs}       % professional-quality tables
\usepackage{amsfonts}       % blackboard math symbols
\usepackage{nicefrac}       % compact symbols for 1/2, etc.
\usepackage{microtype}      % microtypography
\usepackage{lipsum}
\usepackage{fancyhdr}       % header
\usepackage{graphicx}       % graphics
\graphicspath{{media/}}     % organize your images and other figures under media/ folder

\title{Exploring Augmented Reality Games in Accessible Learning: A Systematic Review
}

\author{
  Minghao Cai, Gokce Akcayir, Carrie Demmans Epp\\
  Department of Computing Science \\
  University of Alberta \\
  Edmonton, Canada\\
  \texttt{\{minghao3, akcayir, demmanse\}@ualberta.ca} \\
}

\usepackage{multirow}

\begin{document}
\maketitle

\begin{abstract}
Augmented Reality (AR) learning games, on average, have been shown to have a positive impact on student learning \cite{pellas2019augmenting}. However, the exploration of AR learning games in special education settings, where accessibility is a concern, has not been well explored. Thus, the purpose of this study is to explore the use of AR games in accessible learning applications and to provide a comprehensive understanding of its advantages over traditional learning approaches. In this paper, we present our systematic review of previous studies included in major databases in the past decade. We explored the characteristics of user evaluation, learning effects on students, and features of implemented systems mentioned in the literature. The results showed that AR game applications can promote students’ learning activities from three perspectives: cognitive, affective, and retention. We also found there were still several drawbacks to current AR learning game designs for special needs despite the positive effects associated with AR game use. Based on our findings, we propose potential design strategies for future AR learning games for accessible education.
\end{abstract}

% keywords can be removed

\input{maintext}

% \bibliographystyle{ACM-Reference-Format}
%Bibliography
\bibliographystyle{unsrt}  
\bibliography{reference}

\end{document}

%% file: maintext.tex
\section{Introduction}
Due to the wide adoption of smart mobile devices, such as smartphones and tablets, as well as the availability of Augmented Reality (AR) technology, the past decade has witnessed a rapid growth in the popularity of AR applications in various areas of our lives \cite{parekh2020systematic}. The type of AR interactions that superimpose layers of digital contextualized information over physical settings offer new opportunities for educational experiences. Some studies have found positive effects of AR technology on people’s learning \cite{sirakaya2020augmented,chen2017review}. 

Separate from the investigation of AR in education, many researchers have been studying the use of educational games and have found that this technology can make learning tasks entertaining while providing learning value \cite{dondlinger2007educational}. These studies also showed improved learning when a series of goals, rules, and rewards that are lacking in standard learning approaches were provided through the technology \cite{hanus2015assessing}.

As mentioned above, the positive effects of AR technology and game design in promoting learning performance have been demonstrated. Recently, researchers have started to explore the combination of AR technology and games in education for students who have been designated as requiring adaptation of the learning environment to meet their needs. These students often participate in special-education programs and receive a variety of supports based on their specific needs. 

In a review study examining research on technology use in special education between 2008 and 2012, only one study was found to include AR \cite{liu2013identifying}. In contrast to this study, a recent review that examined technology use in special education between 2014 and 2018 found AR or virtual reality (VR) to be the most frequent multimedia support \cite{olakanmi2020using}. This shows that research on AR use in special education has increased over time. Additionally, AR was found to be an effective learning tool in special education when both group \cite{baragash2020augmentedgroup} and single-subject designs \cite{baragash2020augmented} were used. Similarly, game use in special education has been increasing \cite{cheng2019facilitating} and has been found to be effective \cite{james2020impact}. Despite the demonstrated affordances and increasing prevalence of both AR and game use in special education, we could not find a review study exploring the impact of AR learning games as an assistive learning medium compared to traditional learning tools. To fill this gap, we conducted a systematic literature review to provide a comprehensive understanding of using AR games as accessible learning media and what successful design strategies for such assistive systems would be. The main question is ``what is the current stage of AR learning games in accessible education for students with disabilities considering studies in the past decade?''.

To answer this question, we review studies of existing AR learning game systems for students who have the special education designation. We consider the general characteristics of user studies, effects on student learning performance, and experimental setup including hardware as well as software and the evaluation techniques used. We collected literature from journals and proceedings which are included in major databases, published from 2010 to 2020. This search identified six research articles that presented AR-game-based accessible learning tools. 

Considering the rapid development of AR technology and applications that follow from the continuous upgrading of software and hardware, various definitions of AR have been used. Building on prior work, we describe our specific definition of “AR” and “AR Learning Game” below. 
\begin{itemize}
    \item AR is a technology allowing the superimposition or integration of digital content with physical spaces or objects. It may include the use of technology to provide additional information through the sensory mechanisms of a user. 
    \item An AR Learning Game is an AR technology that incorporates game play.
\end{itemize}
The rest of this paper is structured as follows. Section 2 presents an overview of the methodology and the process of literature search and analysis. Section 3 presents our findings. Section 4 describes our suggestions for designing AR-game-based assistive tools based on our findings. We close the paper with our conclusions.

\section{Methodology}

We adopted the guidelines and steps used in previous systematic review research from related domains \cite{glasziou2001systematic, quintero2019augmented, kitchenham2004procedures} and followed the below steps:
\begin{itemize}
\item {Study planning}
    \begin{itemize}
        \item Formulate research question for the review
        \item Decompose this main question into preliminary categories
    \end{itemize}
\item {Identifying criteria}
    \begin{itemize}
        \item Describe database sources for the literature
        \item Define the inclusion and exclusion criteria
        \item Define search terms
    \end{itemize}
\item {Data collection}
    \begin{itemize}
        \item Preliminary search results
        \item Applying inclusion and exclusion criteria
        \item Data extraction and coding 
    \end{itemize}
\item {Analysis}
\item {Results interpretation and reporting}

\end{itemize}
In the following sections, we describe and report each step. 

\subsection{Study Planning}
To explore the current stage of AR learning games in accessible education for students with disabilities, the following sub-questions were defined:
\begin{itemize}
    \item \textbf{RQ1}: What are the general characteristics (i.e., curriculum subject, learning domain, age group, and learning environment) of the reviewed research? 
    \item \textbf{RQ2}: What are the common kinds of effects of AR learning games for student with disabilities?
    \item \textbf{RQ3}: What kinds of elements or system features are commonly used in AR learning games for student with disabilities?
    \item \textbf{RQ4}: What are the current drawbacks of using AR learning games for students with disabilities?
\end{itemize}

\subsection{Identifying Criteria}

\subsubsection{Database Sources}
To collect as many related journal and proceedings articles as possible, four major interdisciplinary databases were selected as article sources:  Web of Science (WoS) Core Collection, ACM Digital Library, IEEE/IEE Electronic Library (IEL) Online, and Education Resources Information Center (ERIC). WoS, ACM, and IEL are important sources covering peer-reviewed scientific articles \cite{mongeon2016journal}. The ERIC database covers more than 1200 journal titles and is an authoritative collection of journal articles for the field of education \cite{wright2007examining}. These databases are recognized for their coverage and indexing. They are commonly used in literature review research. Thus, they are considered suitable sources for providing a comprehensive article set for our research topic.

\subsubsection{Identifying Criteria}
Three researchers created a codebook that describes the inclusion and exclusion criteria (see below). Two of them participated in the preliminary extraction and coding. 

Initially, we conducted a pilot search where all three read a subset of papers and collaboratively created an initial codebook to use for analyzing the papers. After getting a draft codebook, under the supervision of the third researcher, the two separately analyzed the same small subset of the articles and compared their results to reduce the impact of possible personal bias until the Cohen's kappa inter-coder agreement was 0.91. Once we had decided on the codebook, the remaining papers were split between two researchers, each reading and coding about a half of the remaining papers. 

\subsubsection{Criteria for Literature Inclusion }

We conducted searches between 2010 and 2020 to identify relevant studies in the literature; 2010 was selected as a starting point because using AR for educational purposes gained momentum in 2010 \cite{garzon2019meta}, even though AR technology dates back to 1990’s. So, literature included in our review were studies of educational applications, prototypes, or frameworks that were also AR games between 2010 and 2020. The target population was students with disabilities. More specifically, the below criteria needed to be met for inclusion.
\begin{itemize}
    \item Studies reporting user evaluations or experiments to compare how human learning differs between AR-game and non-AR-game environments.
    \item Studies having outcome measures to indicate the effect of AR-game experience.
    \item Studies reporting the success of the interventions.
    \item Studies that included hypothesis testing to investigate learning differences.
\end{itemize}
Studies were excluded when
\begin{itemize}
    \item papers did not include the term ``Augmented Reality'' or ``AR”''or ``Mixed Reality'' or ``MR'',
    \item pure VR designs were used although the paper claimed to have AR factors,
    \item statistical analyses were not reported, and 
    \item the work was not conducted in an educational context.
\end{itemize}

\subsection{Data Collection}
\begin{figure}[h]
  \centering
  \includegraphics[width=0.8\linewidth]{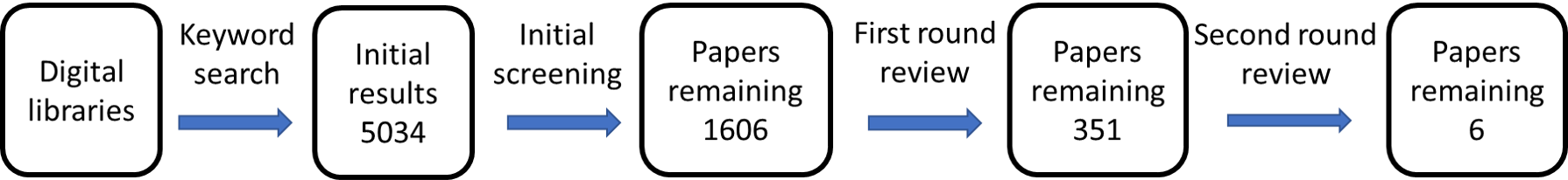}
  \caption{Summary of the analysis process}
  \label{process}
\end{figure}

Figure \ref{process} shows the overview of the approach used to identify the relevant papers. There are three main phases. We found that due to the diversity of diagnoses, some relevant papers would not contain general words about accessible learning such as “universal design” and “accessible” in the title or abstract description. To include as many relevant articles as possible, we did not search for keywords related to games or special education. 

Phase 1: We performed a search in our target databases with keywords based on the criteria above. We used search terms in two categories: (1) AR keywords: ``Augmented Reality'', ``Augmenting Reality'', ``Mixed Reality''; and (2) education keywords: ``learning'', “education”, “training”, “teaching”, “instruction”, “student”, “learner”, “teacher”, “instructor”, “educative”, “teach” ,“learn”. All papers that had more than just an abstract were included, resulting in 5034 total papers. 

Phase 2: We conducted an initial screening to filter articles based on the abstract: 1606 papers remained after this phase. 

Phase 3: We conducted a two-round review based on the full text of the remaining articles. First, a rough scan of the articles was performed. This scan served to eliminate research that did not have a user study or did not meet our definition of AR, and it left 351 papers for further analysis. Following this scan, we reviewed the full text of the article and selected only those papers that fully met the requirements. After completing this process, 6 papers remained: 3 from the Web of Science, 1 from the ACM Library, 1 from IEEE/IEE Electronic Library, and 1 from the Education Resources Information Center. These 6 research papers were included in our analysis.

\subsection{Analysis}

After we had a final set of papers that met our inclusion and exclusion requirements, we read through and coded the papers. This coding focused on measures of application performance, including learning objectives and types of outcome; system features; experimental implementation, including hardware and software; and the assessment approach that the paper detailed in the user studies. See Table \ref{table1} for the codes. 

% Please add the following required packages to your document preamble:
% \usepackage{multirow}
\begin{table}[h]
\caption{Code table for analysis}
\label{table1}
\begin{tabular}{|l|l|l|l|}
\hline
\multicolumn{2}{|l|}{\textbf{Criteria}} &
  \textbf{Description} &
  \textbf{Examples} \\ \hline
\multirow{4}{*}{Learning} &
  Domain &
  The domain   of learning in the user study &
  \begin{tabular}[c]{@{}l@{}}Mathematics, Geometry, Social\\skills\end{tabular} \\ \cline{2-4}   
 &
  Concept &
  The   concept covered in the user study &
  Numeracy   skills \\ \cline{2-4} 
 &
  Effect &
  \begin{tabular}[c]{@{}l@{}}The type of learning effect measured in the \\ study\end{tabular} &
  \begin{tabular}[c]{@{}l@{}}Technical   knowledge, Psychom-\\-otor ability, Motivation\end{tabular} \\ \cline{2-4} 
 &
  Pedagogies &
  How were the concepts learnt in the study? &
  \begin{tabular}[c]{@{}l@{}}Lecture-based,   Project-based, \\ Simulation-based\end{tabular} \\ \hline
Disabilities &
  Type &
  \begin{tabular}[c]{@{}l@{}}Did the system aim to help people with a \\ specific type of disability? If so, what is it?\end{tabular} &
  \begin{tabular}[c]{@{}l@{}}Dyscalculia,   Attention deficit\\ hyperactivity disorder (ADHD)\end{tabular} \\ \hline
\multirow{7}{*}{\begin{tabular}[c]{@{}l@{}}Experimental   \\ Medium\end{tabular}} &
  Platform &
  The   platform of the learning tool &
  Desktop,   Phone, Tablet \\ \cline{2-4} 
 &
  Adaptivity &
  How did   the system support adaptivity? &
  \begin{tabular}[c]{@{}l@{}} Based on   user levels, Based\\on actions\end{tabular} \\ \cline{2-4}
 &
  AR Type &
  The type   of AR used in the system &
  \begin{tabular}[c]{@{}l@{}}Location-based,  Superimposi-\\-tion-based, Recognition-based\end{tabular} \\ \cline{2-4} 
 &
  Interaction &
  The type   of interaction input for users &
  \begin{tabular}[c]{@{}l@{}} Touch,   Gestures, Body move-\\-ments\end{tabular} \\ \cline{2-4}
 &
  Input   sensor &
  \begin{tabular}[c]{@{}l@{}}Did they use a special AR input device or senor \\ to capture users' input? If yes, what was it?\end{tabular} &
  \begin{tabular}[c]{@{}l@{}} Handheld  controller, Depth\\sensor \end{tabular} \\ \cline{2-4}
 &
  \begin{tabular}[c]{@{}l@{}}Registration \\ type\end{tabular} &
  \begin{tabular}[c]{@{}l@{}}The type of registration technique they used to \\ register virtual content in the real world.\end{tabular} &
  Marker-based,   Marker less-based \\ \cline{2-4} 
 &
  Game &
  The type   of game &
  Serious   game, Board game \\ \hline
Assessment &
  Approach &
  \begin{tabular}[c]{@{}l@{}}What kind of approach was used to measure \\ learning?\end{tabular} &
  \begin{tabular}[c]{@{}l@{}} Post-study survey, Scores on \\tests , Completion time\end{tabular} \\ \hline
\end{tabular}
\end{table}

\section{ Findings}
In this section, we describe the results of analyzing the AR learning games that have been used to support accessible learning in educational contexts. We report our detailed answers to the research questions. 

\subsection{The common general characteristics of AR learning games for students with disabilities}

Table \ref{generalchar} shows the distribution of subjects, age groups, learning environments, and the types of disabilities observed across the reviewed user studies. 

% Please add the following required packages to your document preamble:
% \usepackage{multirow}

\begin{table}[h]
\caption{General characteristics}
\label{generalchar}
\begin{tabular}{l|l|l}
\hline
\textbf{Characteristics}      & \textbf{Details}                                & \textbf{The number of papers} \\ \hline
\multirow{2}{*}{Subjects}     & Mathematics                                     & 4                             \\ \cline{2-3} 
                              & Others                                          & 2                             \\ \hline
\multirow{2}{*}{Age group}    & 6-12 years old                                  & 5                             \\ \cline{2-3} 
                              & 12-18 years old                                 & 1                             \\ \hline
\multirow{2}{*}{Environments} & School classroom                                & 4                             \\ \cline{2-3} 
                              & Other indoor places (not specified)             & 2                             \\ \hline
\multirow{5}{*}{Disabilities} & Autism spectrum disorder (ASD)                  & 2                             \\ \cline{2-3} 
                              & Attention deficit hyperactivity disorder (ADHD) & 1                             \\ \cline{2-3} 
                              & Dyscalculia                                      & 1                             \\ \cline{2-3} 
                              & Down syndrome                                   & 1                             \\ \cline{2-3} 
                              & Not specified                                   & 1                             \\ \hline
\end{tabular}
\end{table}

According to the reviewed studies, we found that accessible learning games were most commonly applied to mathematics teaching (67\%), half of which aimed to help students learn knowledge about geometry and the other half aimed to enhance numeracy skills. These mathematical skills are considered to be the basis for meeting the general demands of daily life and learning other domains of knowledge \cite{faragher2005numeracy}. The remaining studies involved training systems designed to enhance core life skills, such as daily life management skills \cite{avila2018towards} and social skills \cite{avila2018design}.

Consistent with these core skill foci, we found that all the research collected was aimed at children and adolescents (<18 years old). Most of the students participating in the experiments were between 6 and 12 years old, and the rest were between 12 and 18 years old. 

With respect to the learning environment, most studies were carried out in school classrooms (67\%). The remaining studies did not specify the experiment environment although most of them were conducted indoors with instructors. 

With respect to the types of disabilities, we found that five of the studies reported specific information about user abilities. Only one study that served students with varied disabilities did not provide specific information. In the reviewed work, Autism Spectrum Disorder (ASD) was the most commonly reported diagnosis. 

\subsection{The common types of learning effect}
In educational application studies, learning effects can be categorized into three domains: the cognitive domain which refers to knowledge and mental skills, the psychomotor domain which refers to motor skills, and the affective domain which refers to feelings and attitudes \cite{hoque2016three}. In addition, we analyzed retention effects since this is a core focus in educational research. We categorized the types of effects of learning into these four main types. The results can be found in Table \ref{learningeffect}. 

\begin{table}[]
\caption{Learning Effect Types}
\label{learningeffect}
\begin{tabular}{l|l|l}
\hline
\textbf{Effect   Type} & \textbf{Samples   effects in search}                    & \textbf{Num} \\ \hline
Cognitive              & Increase   learning gain, Enhance cognitive skills      & 4            \\ \hline
Affective &
  \begin{tabular}[c]{@{}l@{}}Enhance   learning motivation, Enhance willingness to interact, Enhance \\ learning  attention, Enhance engagement, Improve self-confidence,\end{tabular} &
  5 \\ \hline
Psychomotor            & Motor   skills                                          & 0            \\ \hline
Retention              & Learning   gain over time, Motivation over time & 2            \\ \hline
\end{tabular}
\end{table}

\subsubsection{Increased cognitive achievement}
Regarding cognitive learning effects, more than half of the reviewed studies (67\%) reported that AR learning games led to increased learning achievement in terms of content knowledge. The positive effects included enhanced cognitive skills (e.g., improved mathematical reasoning and processing speed \cite{avila2018design}, enhanced problem-solving skills \cite{lin2016augmented}, improved communication skills \cite{lorenzo2019preliminary}) and learning gains \cite{salah2017galaxy}. All these studies reported a positive significant difference except one. In the paper reporting this exception, Lorenzo et al. \cite{lorenzo2019preliminary} describe an AR game that is delivered on mobile phones for enhancing the social skills of children with autism spectrum disorder (ASD). Their preliminary study showed no significant difference in students' general social skill scores between the AR-game and non-AR-game conditions. 

\subsubsection{Affective improvement}
Among the three main domains, affective learning was the most common kind of effect explored when introducing AR learning game applications for students with disabilities. 

Regarding the affective learning effect, most of the reviewed studies explored using AR learning games for stimulating a positive effect on the learners’ attitude towards or experience with the learning content. The detailed effects include attention, engagement, curiosity, enjoyment, self-confidence, and willingness to participating in learning activities. A number of previous studies in education for general users have reported that students tend to have joyful, fun, and playful learning experiences when they play AR games to acquire academic knowledge \cite{lu2015integrating, ayer2016augmented}. A similar result was found when analyzing the reviewed studies of AR learning game applications for learners with disabilities. Most of the studies (83\%) reported that AR learning games led to positive affect, and all of them reported significantly different results when comparing AR learning games to traditional environments. A study by Avila-Pesantez et al. \cite{avila2018towards} described an AR serious game to enhance the daily life functioning of those with Attention Deficit-Hyperactivity Disorder (ADHD). This study reported the system could motivate the emotional competencies and provide continuous interest for problem-solving which could lead to academic success in the long term. Another study for students with Dyscalculia \cite{avila2018design} also showed similar results; the AR game could not only help players maintain continuous interest in mathematical problem-solving tasks but also encouraged students to take an active role in their learning activities. Moreover, in the AR numeracy game for teenagers with Down Syndrome, constructed by Salah et al. \cite{salah2017galaxy}, players tended to have a significantly higher engagement level than those using a regular computer-based setup. In addition to the above effects, AR learning games were also demonstrated to have a positive impact on attention \cite{lorenzo2019preliminary} and concentration \cite{lin2016augmented}. 

\subsubsection{Long-term retention effects}
The reviewed research indicated that AR learning games could help students with disabilities retain what they had learned for a longer time and maintain their interest in learning longer than was seen in non-AR-game learning experiences. 

In our analysis, more than half of the reviewed studies (67\%) measured the outcome results right after the use of an AR learning game. The rest of the studies (33\%) explored relatively long-term impacts on the learners: one study showed a gradual improvement in learning performance and learning motivation in mathematics over a four-week experiment \cite{avila2018design}. The study by Lorenzo et al., reported that the training system led to an affective improvement for children with ASD following a twenty-week intervention \cite{lorenzo2019preliminary}.

\subsection{The common types of measurement}
We further classified the research methods used to measure the learning effects in the reviewed studies into two categories shown in Table \ref{ass}. We did this in the context of the types of effect being measured (see Section 4.2). Because the measurement approaches for cognitive and psychomotor effects are often similar, we merge the two together in this categorization.

\begin{table}[]
\caption{Assessment approach}
\label{ass}
\begin{tabular}{l|l|l}
\hline
\textbf{Effect}                           & \textbf{Approach}      & \textbf{The number of papers} \\ \hline
\multirow{2}{*}{Cognitive \& Psychomotor} & Post-test only & 5                             \\ \cline{2-3} 
                                          & Pre-test and post-test         & 1                             \\ \hline
\multirow{2}{*}{Affective}                & Questionnaire          & 4                             \\ \cline{2-3} 
                                          & Observation            & 2                             \\ \hline
\end{tabular}
\end{table}

According to the reviewed studies, we found that the most commonly used method to measure cognitive and psychomotor effects was “Post-test only” (83\%) which compared student’s performance or knowledge gain with the intervention of AR learning games to those using traditional learning setups such as paper materials and normal computers

For the affective learning effects, our review revealed that the most used main approach was a questionnaire (67\%); the remaining studies used observation. Although the main method of questionnaires was chosen, one of the studies still reported some of the details of student’s behaviors observed by the researchers \cite{singh2019augmented}. It is worth mentioning that except for one study that used an existing user-specific model \cite{avila2018design}, in most of the reviewed studies researchers designed and used their own special-purpose questionnaire.

\subsection{The elements and system features commonly used in AR learning game for special education}
As a relatively complex technology, AR leaning games help to deliver learning content through a medium that is relatively different from traditional learning media (e.g., paper based, physical teaching aids). The beneficial effects discussed in the above sections could come from two aspects of the novel experience (AR and the game) that do not occur in a traditional educational setup. In this subsection, our review sought to explore the different features and elements, which might lead to the different learning outcomes we mentioned above and identify what factors may influence the learning results. 

\subsubsection{Augmented Reality features}
The system features are decomposed into the AR sensing device used to support motion capture or preset markers, AR type used to tie digital content with the physical world, Registration technique used to deploy digital elements into a physical space, and User interaction, which refers to the type of input students use to achieve play. Table \ref{arf} presents information about the use of these different AR features.

\begin{table}[]
\caption{AR factors}
\label{arf}
\begin{tabular}{l|l|l}
\hline
\textbf{System   feature}                 & \textbf{Categories}         & \textbf{The number of papers} \\ \hline
\multirow{3}{*}{AR   sensing device}      & RGB Camera               & 3                             \\ \cline{2-3} 
                                          & Depth sensor                 & 2                             \\ \cline{2-3} 
                                          & Interactive wall            & 1                             \\ \hline
\multirow{3}{*}{User   interaction}       & Gesture / Body movements    & 3                             \\ \cline{2-3} 
                                          & Touch                       & 2                             \\ \cline{2-3} 
                                          & Manipulate physical objects & 1                             \\ \hline
\multirow{2}{*}{AR   type}                & Recognition-based           & 3                             \\ \cline{2-3} 
                                          & Superimposition-based       & 3                             \\ \hline
\multirow{2}{*}{Registration   technique} & Markerless-based                & 5                             \\ \cline{2-3} 
                                          & Marker-based            & 1                             \\ \hline
\end{tabular}
\end{table}

In terms of AR sensing device, we found RGB camera is one of the most commonly used devices (50\%) according to the reviewed studies. This may be because it is easily accessible and relatively inexpensive compared to other sensors on the market. A similar balance was found in a previous review on AR-based application use learning by the general population \cite{fotaris2017systematic}. Regarding the other types of devices, we combined the Depth sensor (33\%) and Interactive wall together because they share a similar function that supports recognition of students’ hand gestures and body movements to realize an intuitive user interface, which makes gesture/body movement (50\%) the most common type of user interaction in the reviewed study. One possible explanation for this difference is a researcher tendency to use the presence of an interactive factor to promote spontaneous interaction which might be inhibited due to students’ abilities. High interaction rates are considered a positive aspect of learning games as they give the learner the sense of freedom and enjoyment that creates a better learning experience without being bored or unchallenged. Also, the types of physical movement that such interaction techniques support can help students with certain types of disabilities enhance their comprehension of cognitive concepts. As previously found \cite{avila2018design}, using Kinect sensors to capture students’ body movements as physical input enhances students’ assimilation of arithmetic concepts, encourages their permanent interest, and maintains highly spontaneous interactions within the learning game. Another study \cite{avila2018towards}, pointed out that providing a natural interface that combined users’ interaction with the objects in a real scenario could stimulate students’ emotional competencies, which would be another benefit to support body movement input in accessible learning.  The Interactive wall design \cite{salah2017galaxy} showed a similar effect where the interactivity increased learning engagement and was associated with increased enjoyment. In addition to the above, touch interaction was commonly used when mobile phones or tablets were used as display devices.  In only one study, the system recognized paper materials and then overlaid instruction information on top of them to assist student manipulation of physical objects. 

\subsubsection{ Game features}
The system features related to Game factors are decomposed into the Platform used to build the learning game, Game type, and Adaptivity that refers to the type of strategies adopted in the game. Table \ref{gamef} presents this analysis.

\begin{table}[]
\caption{Gaming factors}
\label{gamef}
\begin{tabular}{l|l|l}
\hline
\textbf{System   feature}   & \textbf{Categories}             & \textbf{The number of papers} \\ \hline
\multirow{3}{*}{Platform}   & Desktop PC                      & 4                             \\ \cline{2-3} 
                            & Mobile phone/tablet             & 1                             \\ \cline{2-3} 
                            & Projector                       & 1                             \\ \hline
\multirow{3}{*}{Game type}  & Serious game                    & 2                             \\ \cline{2-3} 
                            & Simulation game                 & 2                             \\ \cline{2-3} 
                            & Puzzle game                     & 2                             \\ \hline
\multirow{2}{*}{Adaptivity} & Multiple user levels            & 3                             \\ \cline{2-3} 
                            & Based on user actions/decisions & 4                             \\ \hline
\end{tabular}
\end{table}

For system platform, we found that the most commonly used platform  among the reviewed studies was the desktop PC (67\%), although lightweight mobile AR applications are becoming easily accessible and practical in daily life for enjoyment and demonstration purposes \cite{chatzopoulos2017mobile}. Several possible reasons could explain this finding. First, complicated learning systems usually require powerful computing power, and desktop PCs can have stronger performance than mobile devices. In one study \cite{avila2018design}, when the computer ran a mechanics game with multiple models and calculation rules, it also drove additional AR sensor devices which currently requires more computation than would be available in commercial mobile devices. Second, some emerging sensing devices (e.g., Kinect and interactive project) can only be driven by computers. Third, in a classroom or lab environment, it is easier to find a number of desktop PCs with a unified specification, which makes design and management more convenient for the educator. 

Regarding the application's game type, Serious game, Simulation game, and Puzzle game account for equal proportions (33\%) of the studies. 

Adaptivity is one of the features of game mechanics that can promote learning activities. We found two main types of strategies used according to the reviewed studies. More than a half of the reviewed games (67\%) support multiple paths to achieve learning goals according to learners’ actions or decisions. It is argued that learning occurs when students solve well-ordered problems and subconsciously push themselves to achieve the preset goals \cite{gee2004learning}. Half of the studies provided multiple difficulty levels which change the task and system reactions depending on each player’s skills and capabilities. Such adaptative design is beneficial because students may tend to reach higher enjoyment when the challenge matches their skills, according to the theory of flow proposed by Csikszentmihalyi et al. \cite{csikszentmih2008psychology}. For example, one system would present more complex addition questions when the user level increased \cite{salah2017galaxy}; this was constructed to support “transfer learning” that would help enhance the student's ability to apply the learned knowledge in different situations for those with Down Syndrome.

\subsection{Current drawbacks of using AR learning games for student with disabilities}

In this subsection, we discuss the limitations and challenges of AR learning games for accessible leaning based on the observations from the analyzed articles. 

First, AR learning game systems that support more complex functions often require a desktop computer platform and high-end hardware to support powerful computing. Connection to additional devices including sensors and display devices would be needed for providing multi-modal interaction in learning. Thus, at this stage, a complete set of equipment will be relatively expensive. A need for more affordable solutions poses challenges to popularization. In the meantime, the large and cumbersome equipment reduces portability and limits the promotion of these approaches to other areas in the learner’s life. As a result, many such systems remain at the experimental stage and can only be used in specific research environments, even if the system provides excellent learning aids. This limitation could be reduced by the rapid development of lightweight and high-performance computing equipment and sensing equipment.

Second, the AR-game learning method brings new challenges to teachers. As mentioned in some of the studies \cite{lin2016augmented, singh2019augmented}, the introduction of new technology requires curriculum designers to acquire specific computer knowledge. At the same time, the lecturer also needs more time to master certain knowledge to ensure the smooth use of the system when delivering learning content. Potential solutions include optimizing the interactive interface or forming a template of skill sets to help teachers master basic technical skills in advance.

Third, the diversity and particularity of the needs of students with disabilities makes system design difficult. Design elements that are common in other educational settings may pose challenges when used with certain groups of people. As one study involving students on the autism spectrum shows \cite{singh2019augmented}, technologies requiring learner autonomy may not work because of the often-cited dependence of these students on their teachers and the important role that their relationship plays in learning. To solve this problem, we may need more intelligent AR technologies that can better accommodate variable needs in learning. Investigating and summarizing technical models suitable for certain patterns of needs can also improve design efficiency.

Fourth, in the reviewed studies, most of the participants had no relevant experience with AR or similar technologies before participating in the experiment. This suggests we cannot rule out a potential novelty effect of introducing a new technology, especially when considering affective outcomes. More research should be done focusing on exploring the potential correlation.

%Fifth, differences based on gender were not investigated. In the collected studies, some report the gender of students but most did not take it into account during sampling or analysis. This might be due to the small sample size and warrants additional attention.

\section{Suggestions for Design Strategy}

In this section, we present our main suggestions on the design of AR games in accessible learning applications based on our previous discussion.

\subsection{Clarify the Learning Effect}
From our analysis, the impact of AR games on special education is multifaceted, including various aspects from cognitive acceleration to affective enhancement. Corresponding to different design goals, there will be different system characteristics. For example, supporting three-dimensional virtual display can help students understand complex geometric content and supporting natural interaction methods, such as gesture input, often has good results when promoting student participation. Therefore, when designing the system, the designer should first clarify the type of learning effect that is desired in order to select the corresponding design elements.

\subsection{Understand Individualized Needs}
In our reviewed work, most of the systems were oriented towards an isolated group of students with a specific type of disability. This makes sense from a research perspective but is not always useful if we want systems to be integrated into classrooms. Many special education classrooms have a mixture of students, each with their own differentiated abilities and supports. Since the system design may affect different groups of learners in different ways, it is important to consider who will and will not benefit from that design. It is also challenging to find a truly diverse design that will be effective for all students. Therefore, designers need to consider the types of approaches that they will use if they are to meet the specific needs of target users. Given the highly differentiated needs of these students it may be advisable to support configuration and adaptivtiy based on student abilities.  

\subsection{Differentiate the Impact of System Characteristics}
From our findings, we can see that the diverse combinations of AR and Game technologies contribute to the observed diversity of system functions and features. From the perspective of AR, for example, designs can be divided into marker-based and markerless-based and can also be divided into superimposition-based and recognition-based. From the perspective of game mechanics, they can be divided into simulation games, strategy games, and a variety of other types of games. The existing designs often contained more than one system feature, but there are very few studies investigating the learning impact of a single feature. Therefore, designers should consider and study the multi-faceted influence of one element and the joint effect of multiple different elements on student's learning when building the system.

\section{Limitations}
Although our literature review gave us an understanding of the use of AR games for current accessible education research, we were unable to conduct further quantitative analysis due to the limited number of papers that were of sufficient quality to enable meta-analysis. Considering the four major databases that were searched and the 10-year period we used, our search can be described as comprehensive. So, a lack of high-quality research that focuses on the outcomes of AR game use in special education is probably the issue. To fully understand the design approaches used independent of their effectiveness, relevant products that are publicly available can be examined. Such analyses would then only include design attributes and not their influence on student outcomes. Given the current state of work in this area, the results of our analysis provide constructive guidance for future research and design.

\section{Conclusion}
In this paper, we conducted a systematic review of previous studies to provide a comprehensive understanding of the use of AR games in accessible learning. Our analysis summarizes common experimental elements and system characteristics mentioned in the literature and further discusses the multi-faceted impact of AR games. The findings indicate that AR learning games could be implemented to help students with special needs by increasing their learning abilities and improving their learning performance for different learning purposes. It was also found that using AR learning games can positively impact student affect during learning activities. Furthermore, our review reveals several limitations and challenges that are associated with current designs. Finally, based on our findings, we suggest three design strategies for future studies that could help us maximize the positive effects of AR learning games. In light of these findings, we can say that AR game applications hold potential for supporting the specific needs of learners. However, considering the special education context and the variability in learners’ needs, it is known that one size does not fit all. For this reason, there is a need for a vast number of studies exploring the design, implementation, and evaluation of AR games for learners with differentiated needs.

%%
%% The next two lines define the bibliography style to be used, and
%% the bibliography file.

%%

%%
%% End of file `sample-manuscript.tex'.